\documentclass[fleqn,twoside]{article}
\usepackage{espcrc2-hepph0502062}
\usepackage{latexsym,amssymb,amsmath,epsfig}  %%FRK

\usepackage{graphicx}
\renewcommand{\text}{\mathrm}   %%for proceedings, not for transp
\newcommand{\thetamass}{\underline{\theta}} %%mass parameters must be underlined
\newcommand{\bdi}{\begin{displaymath}}
\newcommand{\edi}{\end{displaymath}}
\newcommand{\bfi}{\begin{figure}}
\newcommand{\efi}{\end{figure}}

\newcommand{\beq}{\begin{equation}}
\newcommand{\eeq}{\end{equation}}
\newcommand{\beqa}{\begin{eqnarray}}
\newcommand{\eeqa}{\end{eqnarray}}

\newcommand {\SM}     {Standard Model}
\newcommand {\gsim}{\mathrel{\hbox{\rlap{\lower.55ex \hbox {$\sim$}}
            \kern-.3em \raise.4ex \hbox{$>$}}}}
\newcommand {\lsim}{\mathrel{\hbox{\rlap{\lower.55ex \hbox {$\sim$}}
            \kern-.3em \raise.4ex \hbox{$<$}}}}
\title{Lorentz and CPT violation: a simple neutrino-oscillation model}

\author{Frans R. Klinkhamer\address{Institute for Theoretical Physics,
University of Karlsruhe (TH), 76128 Karlsruhe, Germany}}

\begin{document}

\begin{abstract}
We present a three-parameter model for
three flavors of massless left-handed neutrinos with
Fermi-point splitting and tri-maximal mixing angles.
One of these parameters is the T--violating phase $\delta$,
for which the experimental results from K2K and KamLAND appear
to favor a nonzero value.
Future experiments, in particular MINOS,
will be able to test this simple model.
Possible implications for neutrino factories are briefly discussed.
\vspace{1pc}
\end{abstract}

\maketitle
\mathindent=0pt

\section{INTRODUCTION}

The experimental results on neutrino oscillations
(as discussed by other talks at this Conference) suggest
the following pattern of mixing angles:
\beq
\thetamass_{\,21}  \approx \thetamass_{\,32}  \approx \pi/4   \,,\quad
\thetamass_{\,13}  \approx  0    \,,
\label{MassThetapattern}
\eeq
assuming the validity of the mass-difference mechanism
\cite{BilenkyPontecorvo1978}.
But, perhaps, one angle need not be small if
there are other mechanisms operative?

In fact, there is the possibility that
Lorentz invariance is  not a fundamental symmetry but an emergent
phenomenon \cite{{VolovikBook}}.
Massless (or nearly massless) neutrinos could then provide us with a
window on ``really new physics.''

In this talk, we discuss an idea from condensed-matter physics,
namely Fermi point splitting by a quantum phase transition
\cite{KlinkhamerVolovikJETPL,KlinkhamerVolovikIJMPHA}.
%namely the splitting of a multiply degenerate Fermi point by a quantum phase
%transition \cite{KlinkhamerVolovikJETPL,KlinkhamerVolovikIJMPHA}.
(Fermi points ${\bf p}_n$ are points in three-dimensional momentum space at
which the energy spectrum of the fermionic quasi-particle has a zero.)
The neutrino-os\-cil\-la\-tion model
considered \cite{KlinkhamerJETPL,KlinkhamerNeutrinoModel}
is the simplest one possible
with \emph{all} mixing angles large and mass differences vanishing
\emph{exactly}. Needless to say, this model may be only a first approximation.

Note that the idea of neutrino oscillations from
Fermi-point splittings is orthogonal to the
suggestion of having CPT--violating masses to explain LSND
(see, e.g., Ref.~\cite{BarenboimLykken} and references therein).

\section{FERMI-POINT-SPLITTING ANSATZ}

In the limit of vanishing Yukawa couplings,
the chiral fermions of the Standard Model
may still have Fermi-point splittings in their dispersion law,
\beq
\bigl( E_{a,f}({\bf p}) \bigr)^2  =
\Bigl(\,c\, |{\bf p}| + b_{0a}^{(f)} \Bigr)^2 \,.
\label{SMdispLaw-timelike}
\eeq
Here, $a$ labels the 16 types of massless left-handed Weyl fermions
in the \SM~ (with a hypothetical left-handed antineutrino
included) and $f$ distinguishes the 3 known fermion families.

An example of Fermi-point splitting is given by
the following factorized \emph{Ansatz}
\cite{KlinkhamerVolovikIJMPHA}:
\begin{equation}
b_{0a}^{(f)} =  Y_a \; b_{0}^{(f)}\,,
\label{SMb0pattern}
\end{equation}
with $Y_a$ the \SM~hypercharges of the left-handed fermions.
For this special pattern, the induced electromagnetic Chern--Simons term
cancels out exactly. This allows for $b_{0}$ values very much larger than
the experimental upper limit on the Chern--Simons energy scale,
which is of the order of $10^{-33}\;{\rm eV}$ \cite{CarollFieldJackiw}.

Independent of the particular pattern of Fermi-point splitting,
the dispersion law of a massless
left-handed neutrino can be written as
\beq
\bigl( E_{\nu_L,f}({\bf p}) \bigr)^2  =
\Bigl(\,c\, |{\bf p}| - b_0^{(f)}\, \Bigr)^2  \,.
\label{DispLaw-nu}
\eeq
The right-handed antineutrino
is assumed to have the same dispersion law as (\ref{DispLaw-nu})
but with a plus sign in front of $ b_0^{(f)}$
(the case with a minus sign is also discussed in
Ref.~\cite{KlinkhamerNeutrinoModel}).

More generally, one may consider for large momenta $|{\bf p}|$:
\begin{equation}
 E({\bf p})       \sim
c\, |{\bf p}| \pm b_0 +  \frac{m^2\,c^4}{2 |{\bf p}| c}
+ \text{O}\left(|{\bf p}|^{-2}\right)\,.
\label{DispLaw-b0m}
\end{equation}
The energy change from a nonzero $b_{0}$ always dominates the
effect from $mc^2$ for large enough $|{\bf p}|$.
In order to search for Fermi-point splitting, it is therefore
preferable to use neutrino beams with the highest possible energy.

In this talk, we set all neutrino masses to zero. Let us emphasize that
this is only a simplifying assumption and that there may very well
be additional mass terms, as in Eq.~(\ref{DispLaw-b0m}).
However, with both mass terms and Fermi-point splittings present,
there is a multitude of mixing angles and phases to consider,
which is the reason to leave the masses out in an exploratory
analysis.

\section{THREE-PARAMETER MODEL}

The flavor states $|A\rangle, |B\rangle, |C\rangle$
and the left-handed propagation states $|1\rangle, |2\rangle, |3\rangle$
with dispersion law (\ref{DispLaw-nu}) are related by
a unitary  $3 \times 3$ matrix $U$:
\beq
\left( \begin{array}{c}|A\rangle\\ |B\rangle\\ |C\rangle \end{array} \right)
  = \; U \;
\left( \begin{array}{c}|1\rangle\\ |2\rangle\\ |3\rangle  \end{array} \right)
   \,.
\label{ABC=U123}
\eeq
The standard parametrization of $U$ has
one
phase $\delta  \in [0,2\pi)$,
and three
mixing angles $\theta_{21}, \theta_{32}, \theta_{13}  \in [0,\pi/2]$.

In order to emphasize the difference with the
current paradigm (\ref{MassThetapattern})
for the mixing angles $\thetamass_{\,ij}$ associated with mass terms,
we take the mixing angles from (\ref{ABC=U123}) to be \emph{tri-maximal}:
\begin{subequations} \label{thetapattern}
\begin{eqnarray}
\theta_{21}  &=& \theta_{32}= \arctan 1=\pi/4  \,, \\
\theta_{13}  &=& \arctan \sqrt{1/2} \approx \pi/5 \,.
\end{eqnarray}
\end{subequations}
These particular values maximize,
for given phase $\delta$, the T--violation (CP--non\-con\-ser\-va\-tion)
measure $J \equiv \frac{1}{8}\, \cos\theta_{13}\,\sin 2\theta_{13}\,
            \sin2\theta_{21}\,\sin 2\theta_{32}\, \sin\delta$
of Ref.~\cite{Jarlskog}.

The Fermi-point-splitting energies $b_0^{(f)}$ are assumed to be
positive and to increase with $f$, giving rise to two positive parameters:
\begin{subequations}
\begin{eqnarray}
B_0 &=& b_0^{(2)}- b_0^{(1)} \,,
\label{parameter-B0}
\\[2mm]
r &=&  \bigl(\,b_0^{(3)}- b_0^{(2)} \,\bigr)/
       \bigl(\,b_0^{(2)}- b_0^{(1)}\,\bigr)\,.
\label{parameter-r}
\end{eqnarray}
\end{subequations}
All in all, the model \cite{KlinkhamerNeutrinoModel}
has three parameters:
%All in all, there are three parameters in the model:
\begin{itemize}
\item
the basic energy-difference scale $B_0$,
\item
the ratio $r$ of the two energy steps $\Delta b_0$,
\item
the T--violating phase $\delta$.
\end{itemize}
This model  will be called
the ``simple'' Fermi-point-splitting
model in the following (a more general Fermi-point-splitting model
would have arbitrary mixing angles  $\theta_{ij}$).

\section{OSCILLATION PROBABILITIES}

For large enough neutrino energy (that is,
\mbox{$E_\nu \geq \text{max}\;[\,b_0^{(f)}\,]\,$}),
the tri-maximal model gives neutrino oscillation probabilities
\beq
P(X \rightarrow Y)\,,\quad \text{for}\;\; X,Y \in \{A,B,C\},
\eeq
in terms of the dimensionless distance
\beq
l \equiv B_0\, L /(h\,c)
\label{l}
\eeq
and the other two model parameters, $r$ and $\delta$.
With the assumed dispersion laws,
the same probabilities hold for antineutrinos,
\beq
P(\bar{X} \rightarrow \bar{Y}) = P(X \rightarrow Y) \,.
\label{probabilitiesCP}
\eeq

For the model probabilities, the time-reversal asymmetry
between $A$--type and $C$--type neutrinos is given by
\beq
a^\text{(T)}_{CA} \equiv
\frac{P(A\rightarrow C)-P(C\rightarrow A)}
     {P(A\rightarrow C)+P(C\rightarrow A)} \propto \sin\delta\,,
\label{asymmT}
\eeq
whereas the CP--asymmetry vanishes identically.

\begin{table*}[t]
\caption{Lengths $[\text{km}]$ and time-reversal
asymmetries $a^\text{(T)}_{\mu e}$
for selected model parameters $B_0\,[10^{-12}\,\text{eV}]$ and $r$, \newline
with phase $\delta$ $=$ $\pi/4 \pmod{\pi}$
and identifications (\ref{ABCidentification}ab).}
\newcommand{\m}{\hphantom{$-$}}
\newcommand{\cc}[1]{\multicolumn{1}{c}{#1}}
\renewcommand{\tabcolsep}{1.25pc} % enlarge column spacing
\renewcommand{\arraystretch}{1.125} % enlarge line spacing
\begin{tabular}{@{}ll|lllll}
\hline
$B_0\phantom{\stackrel{+}{T}}$
&$r$
&$\lambda$
&$L_\text{magic}$
&$a^\text{(T)}_{\mu e}\,(L_\text{magic})$
&$L^\prime\phantom{\!}_\text{magic}$
&$a^\text{(T)}_{\mu e}\,(L^\prime\phantom{\!}_\text{magic})$\\[0.1mm]
\hline
0.72&1          &1724 &386&$+89\,\%$ &1338 &$-89\,\%$\\
1.04&1/2        &2381 &357&$+99\,\%$ &2024 &$-99\,\%$\\
1.29&1/4        &3846 &395&$+81\,\%$ &3451 &$-81\,\%$\\
\hline
\end{tabular}
\label{tableLengths}
\end{table*}

\section{PARAMETERS AND PREDICTIONS}

\subsection{General predictions}

Two general predictions \cite{KlinkhamerJETPL} of the Fermi-point-splitting
mechanism of neutrino oscillations are:
\begin{itemize}
\item
\emph{undistorted} energy
spectra for the reconstructed $\nu_\mu$ energies in, for example,
the current K2K experiment
and the future MINOS experiment;
\item
survival probabilities close to 1
for \emph{all} reactor experiments at $L \approx 1\,\text{km}$
(e.g., CHOOZ and double-CHOOZ), at least up to an accuracy of order
$(\Delta b_0\, L/\hbar c)^2 \approx   10^{-4}$
for $\Delta b_0  \approx 2 \times 10^{-12}\;{\rm eV}$
(see below).
\end{itemize}
Both predictions hold only if mass-difference effects from
the generalized dispersion law (\ref{DispLaw-b0m}) are negligible
compared to Fermi-point-splitting effects.
With (anti)neutrino energies in the $\text{MeV}$  or $\text{GeV}$
range, this
corresponds to $\Delta m^2  \lesssim 10^{-6}\,\text{eV}^2/c^4$
or  $\Delta m^2  \lesssim 10^{-3}\,\text{eV}^2/c^4$,
respectively.

\subsection{Preliminary parameter values}

Comparison of the model probabilities
and the \emph{combined} K2K and KamLAND results gives the following
``central values'' \cite{KlinkhamerNeutrinoModel}:
\begin{subequations}\label{parameters}
\begin{eqnarray}
B_0    &\approx& 10^{-12}\;{\rm eV} \,, \\
r      &\approx& 1/2 \,,                \\
\delta &\approx& \pi/4 \;\;\text{or} \;\; 5\pi/4 \,,
\end{eqnarray}
\end{subequations}
with identifications
\begin{subequations}
\label{ABCidentification}
\begin{align}
\bigl( |A\rangle\,,  |B\rangle\,,  |C\rangle \bigr) &=
\bigl( |\nu_e\rangle \,, |\nu_\tau\rangle \,,|\nu_\mu\rangle \bigr)
\Big|^{\,\delta \approx \pi/4}\;,
\label{ABCidentificationQuarterPi}\\
\bigl( |A\rangle\,,  |B\rangle\,,  |C\rangle \bigr) &=
\bigl( |\nu_e\rangle \,, |\nu_\mu\rangle \,,|\nu_\tau\rangle \bigr)
\Big|^{\,\delta \approx 5\pi/4}.
\label{ABCidentificationFiveQuarterPi}
\end{align}
\end{subequations}
Detection of an interaction event (e.g., $\mu^{-}$ decay) is needed to decide
between the options (\ref{ABCidentification}ab).

\subsection{Specific predictions}

Detailed model predictions can be found in
Ref.~\cite{KlinkhamerNeutrinoModel}, in particular
figures relevant to MINOS and T2K (JPARC--SK).
If the model has any validity, MINOS should be able
to reduce the range of $r$ values
compared to the range allowed by the current K2K data.

Table \ref{tableLengths} gives the wavelength $\lambda$,
the distance $L_\text{magic}$  which maximizes
the time-reversal asymmetry
$a^\text{(T)}_{\mu e}$ from Eq.~(\ref{asymmT}),
and the other magic distance
$L^\prime\phantom{\!}_\text{magic}$
which minimizes this T--asymmetry.
Observe that $\,L_\text{T2K} = 295 \; \text{km}$ is of the same
order of magnitude as $L_\text{magic}\,$, which would make having both
$\nu_e$ and $\nu_\mu$ beams from JPARC especially interesting.

\section{OUTLOOK}

We propose to use the following checklist:
\begin{itemize}
\item
\emph{equal survival probabilities
$P(\nu_\mu \rightarrow \nu_\mu)$ for the low- and high-energy beams of
MINOS?}
\item
\emph{appearance probability
\mbox{$P(\nu_\mu \rightarrow \nu_e)$} from MINOS \mbox{above a few percent?}}
\item
\emph{consistent fit of the (simple) Fermi-point-splitting model to the
combined data from K2K, MINOS, and ICARUS/OPERA?}
\end{itemize}
If this  more or less works out, one would have to reconsider the
future options based on the relevant \emph{energy-independent} length scales
of the (simple) Fermi-point-splitting model
[cf. Table \ref{tableLengths}]
or those of an extended version
with additional mass terms ($\Delta m^2 \approx 10^{-4}\,\text{eV}^2/c^4\,$?).
These future options include  beta beams and neutrino factories.
But, first, let's see what MINOS finds $\ldots$

\end{document}